\documentclass{desyproc}

\begin{document}
%------------------------------------
\title{Experimental Summary}

%for single authors the superscripts are optional
\author{{\slshape Peter Bussey}\\[1ex]
Deparftment of Physics and Astronomy, Faculty of Physical Sciences,\\
University of Glasgow, Glasgow G12 8QQ, U.K.}

% if the proceedings are available online (e.g. at Indico)
% please enter the contribution ID or file_name below for the DOI
\contribID{20}
%\contribID{smith\_joe}

% TO THE CONFERENCE EDITORS: 
% please update the following information      
% before sending the template to the authors
\confID{1407}  % if the conference is on Indico uncomment this line
\desyproc{DESY-PROC-2009-03}
\acronym{PHOTON09} % if you want the Acronym in the page footer uncomment this line
\doi  % if there is an online version we will register DOIs

\maketitle

\begin{abstract}
I summarise the content of the experimental talks of the conference.
\end{abstract}

\section{Introduction}

In this account, I am attempting to summarise a substantial collection
of very diverse experimental talks, all of which carry some kind of
association with the physics of photons.  This was of course a particle
physics conference, and so we are dealing with photons that in some
sense behave as elementary particles.  However
this does not necessary mean that the photons treated in a given context
were always of high energy.  It is part of the richness of the subject
that the photons in the experiments presented here
could vary in energy by orders of magnitude, and yet still maintain
the connection with elementary particle physics. I intend to depart from
the presentational order of the talks and start with those that
involved photons of lowest energy, finishing with those of the
highest.  Much material has had to be omitted, but it can be found in
the respective single-topic talks. These should in any case be
consulted for more details and for references to the published work.
The following sections, therefore, are very much an ``invitation to
further reading''.

\begin{figure}[b]
\centerline{
\includegraphics[bbllx=0,bblly=0,bburx=666,bbury=117,width=0.65\textwidth]{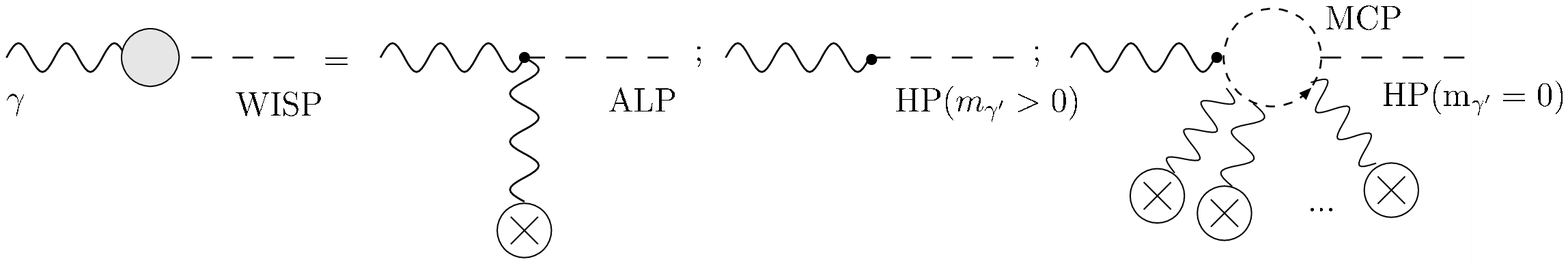}
\hspace*{0.05\textwidth}
\raisebox{0.05\textwidth}{\includegraphics[width=0.3\textwidth]{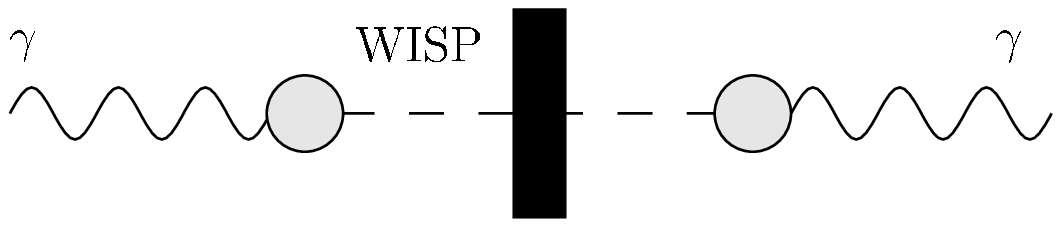}}
}
\caption{Schematic production and detection of WISPS (Lindner).}
\label{Fig:axionsa}
\end{figure}

\begin{figure}[t]
\centerline{
\includegraphics[width=0.6\textwidth]{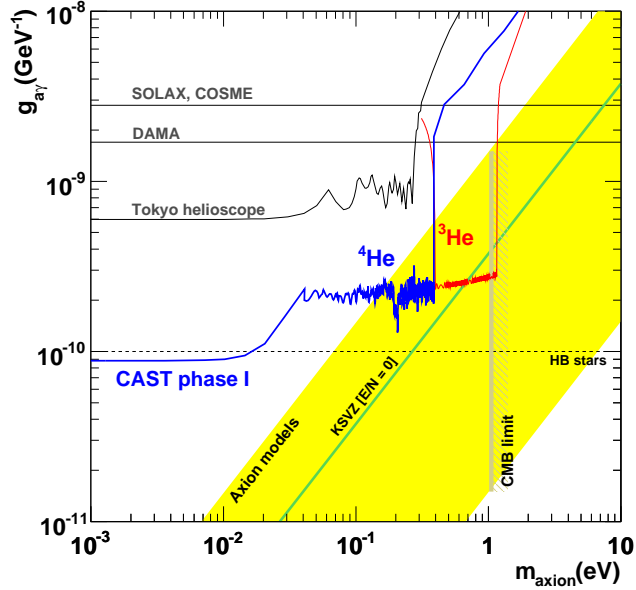}
}
\caption{Present limits
from ALP experiments (Cantatore).
}
\label{Fig:axionsb}
\end{figure}

\section{Axions and their relatives}
Axions and their relatives constitute a wide class of hypothetical
neutral particles that couple to photons.  They have been proposed in
different contexts, and in his talk Joerg Jaeckel presented the
motivations for looking for these various objects.  There is a
theoretical problem with explaining why CP is conserved in QCD, the
so-called ``strong CP problem'', since the theory's vacuum structure
permits a CP violation.  The axion is a proposed particle whose
presence prevents this from happening. It must be very light and very
weakly interacting, but it should couple to two photons. It is an
example of a more generic class of ``WISPs'' -- Weakly Interacting
Sub-eV Particles.  In the case of axion searches, one approach is to
use a strong magnetic field to supply a virtual photon, with which
photons from a strong laser beam interact, hopefully generating some
axions. A barrier then absorbs the remaining laser beam and everything
else except the axions, which proceed through the barrier into a
further region of magnetic field, where they regenerate a photon that
may be observed.  Thus ``light shines through a wall''
(Fig.~\ref{Fig:axionsa}).

Giovanni Cantatore provided a comprehensive review of a diversity of
attempts to discover WISPs: in general the search is for ALPs,
``Axion-Like Particles''.  The lowest energy photons featuring in this
conference were those of the ADMX collaboration, in which cosmological
relic axions are invited to interact with microwaves in a
cavity. Within a relatively narrow band of axion masses in the
micro-eV region, this experiment is currently unique in actually reaching the
sensitivity of the theories; however no signal was found.  The CERN
axion search CAST points a powerful magnet at the sun to detect ALPS
in the meV energy range, as does the Tokyo Helioscope
(Fig.~\ref{Fig:axionsb}).  PVLAS approaches the problem by using an intense laser beam to look for
an effective vacuum dichroism in a magnetic field.  Its former claimed
result has now been disconfirmed. So no signal has been found yet, but
the searches go on since the sensitivity in most cases still needs to
be improved substantially.  Axel Lindner described a far-ranging
collection of further proposed theoretical end experimental ideas for
improved ALP and WISP searches of various kinds.  There are
hypothetical ``Mini-Charged Particles'' and even a class of ``Hidden
Photons'' that do not need a magnetic field but will just appear in a
vacuum tube! Clearly this area is proving an immensely fertile ground
for theoretical imagination as well as experimental ingenuity.

\begin{figure}[t]
\centerline{
\includegraphics[bbllx=300,bblly=306,bburx=577,bbury=678,width=0.38\textwidth,angle=270,clip=true]{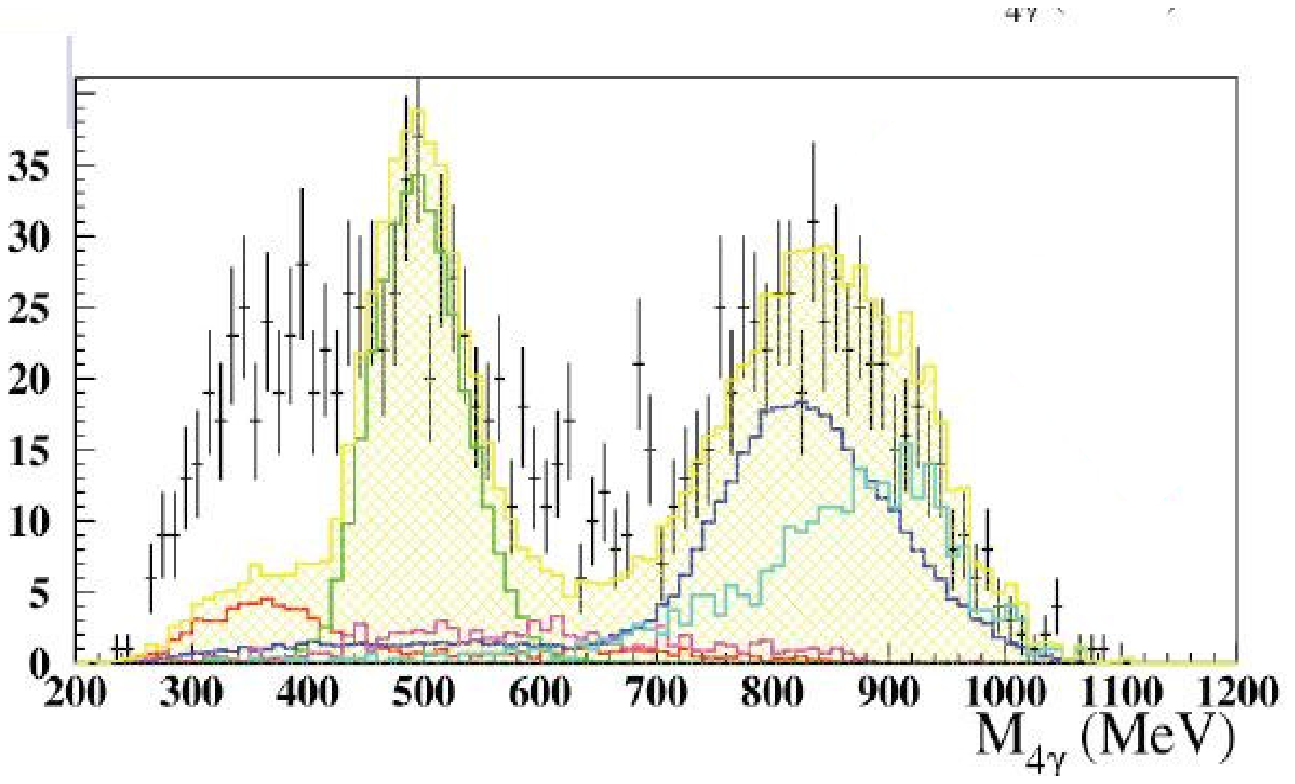}
\hspace*{0.01\textwidth}
\raisebox{0.075\textwidth}{
\includegraphics[bbllx=145,bblly=81,bburx=500,bbury=509,width=0.38\textwidth,angle=270,clip=true]{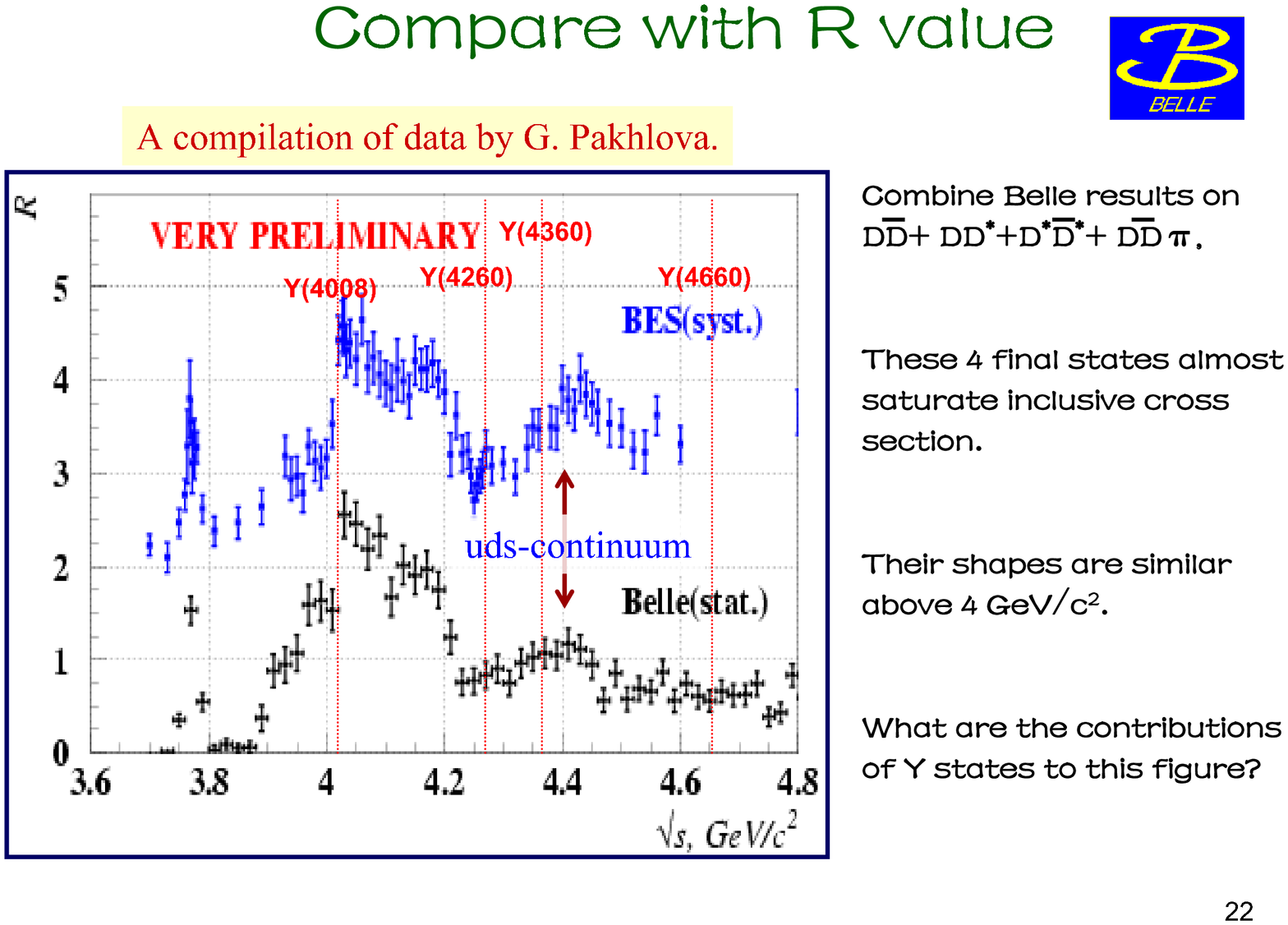}
}}
\vspace*{-0.05\textwidth}
\caption{(a) 4-photon mass spectrum showing low-mass excess 
at KLOE (Di Donato); 
(b) $R$ ratio of hadrons to muon pairs in $e^+e^-$ collider data
showing the charm threshold and resonances (Wang).  }
\label{Fig:elpos}
\end{figure}

\section{Photons at electron-positron colliders}

The electron-positron colliders from which results were presented at this
conference were DA$\Phi$NE, BaBar and Belle.  Data taking at
DA$\Phi$NE ended in 2006, with 2.5 fb$^{-1}$ of data at the $\phi$
mass, and BaBar have finished with 553 fb$^{-1}$ of data. Many results
on radiative $\phi$ decays and photon-photon processes are now
becoming available, a selection of which was presented by Camilla Di
Donato. The radiative $\phi$ decays that were studied include those in which a
single light meson is accompanied by a photon, and those where two mesons are
produced. There are open questions here regarding the existence and
properties of scalar mesons below 1 GeV in mass, which these data are
uniquely posed to answer. In particular, there is the
long-standing question of whether a $\sigma(600)$ meson
exists. Production of a $2\pi^0$ final state shows an
anomaly that could indicate the presence of a new effect in this
channel, but apparently at a lower energy than 600 GeV
(Fig.~\ref{Fig:elpos}(a)).  A gluonium content within the $\eta'$ is
indicated, and analyses have been started on the decay of the $\eta$
and $\eta'$ into $\pi^+\pi^-\gamma$.

Initial-state photon radiation is able to make a number of resonant
states available for study at the $B$ factories.  A wide range of
masses is scanned in this way, and the high statistics available are
generating some interesting results. Xiao Long Wang showed how exotic
charm structures referred to as $Y(4008)$, $Y(4260)$, $Y(4360)$,
$Y(4660)$ are now under study. They are formed from
charmonium plus hadron pairs and are not yet all fully understood -- a
topic of considerable interest in the context of the quark model of mesons
(Fig.~\ref{Fig:elpos}(b)).  It is interesting that when two charm
mesons are observed in the final state, no evidence of the $Y$ states
is seen. There is also the $Y(2175)$, which may be an excited strange quark
state, and its presence has been confirmed at BES. Here are some
ongoing investigations where Belle is best placed to give further
answers.

\begin{figure}[t]
\vspace*{-0.04\textwidth}
\centerline{
\includegraphics*[width=.42\textwidth]{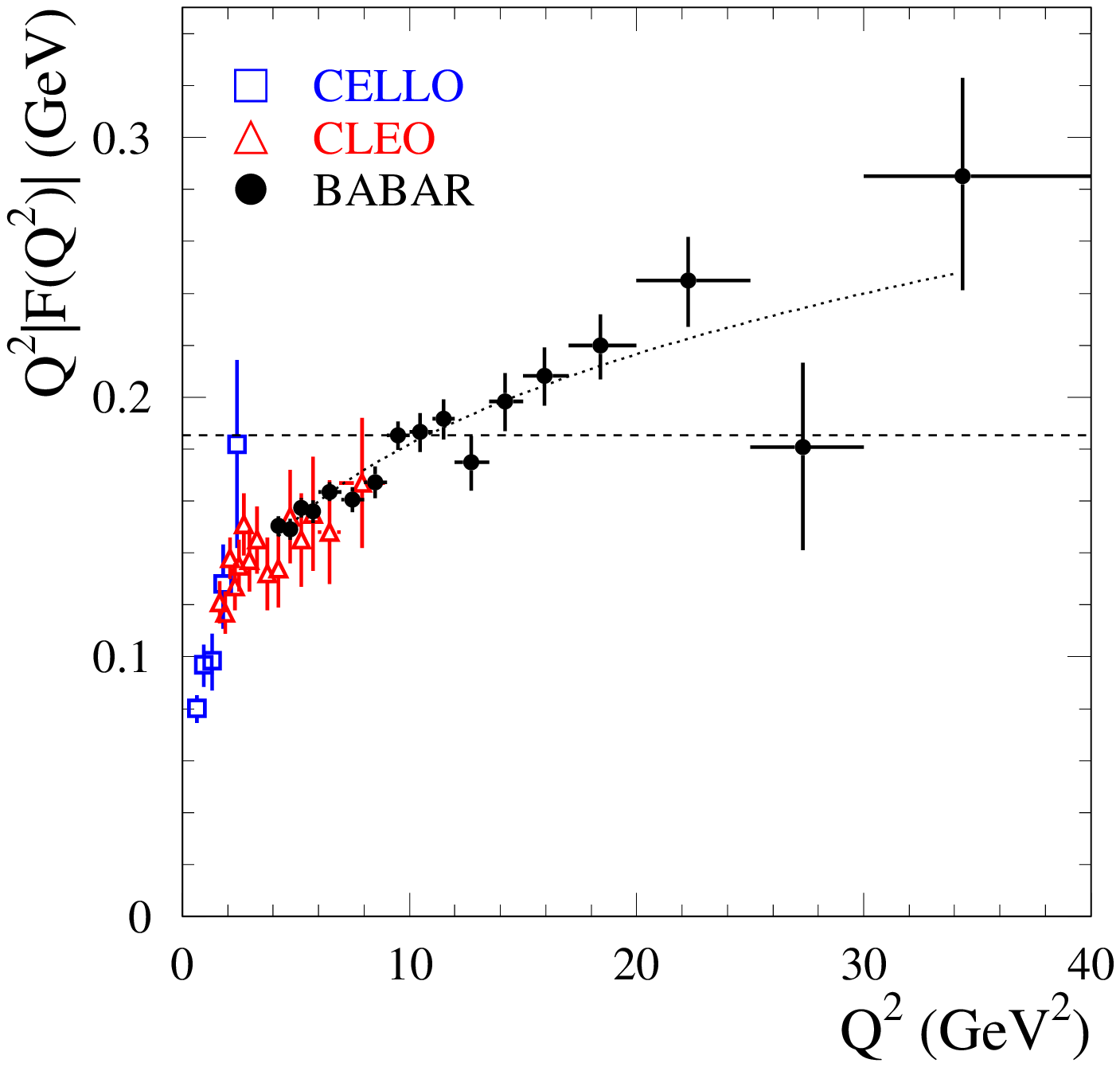}
\hspace*{0.05\textwidth}
\raisebox{0.033\textwidth}{
\includegraphics[width=.53\textwidth]{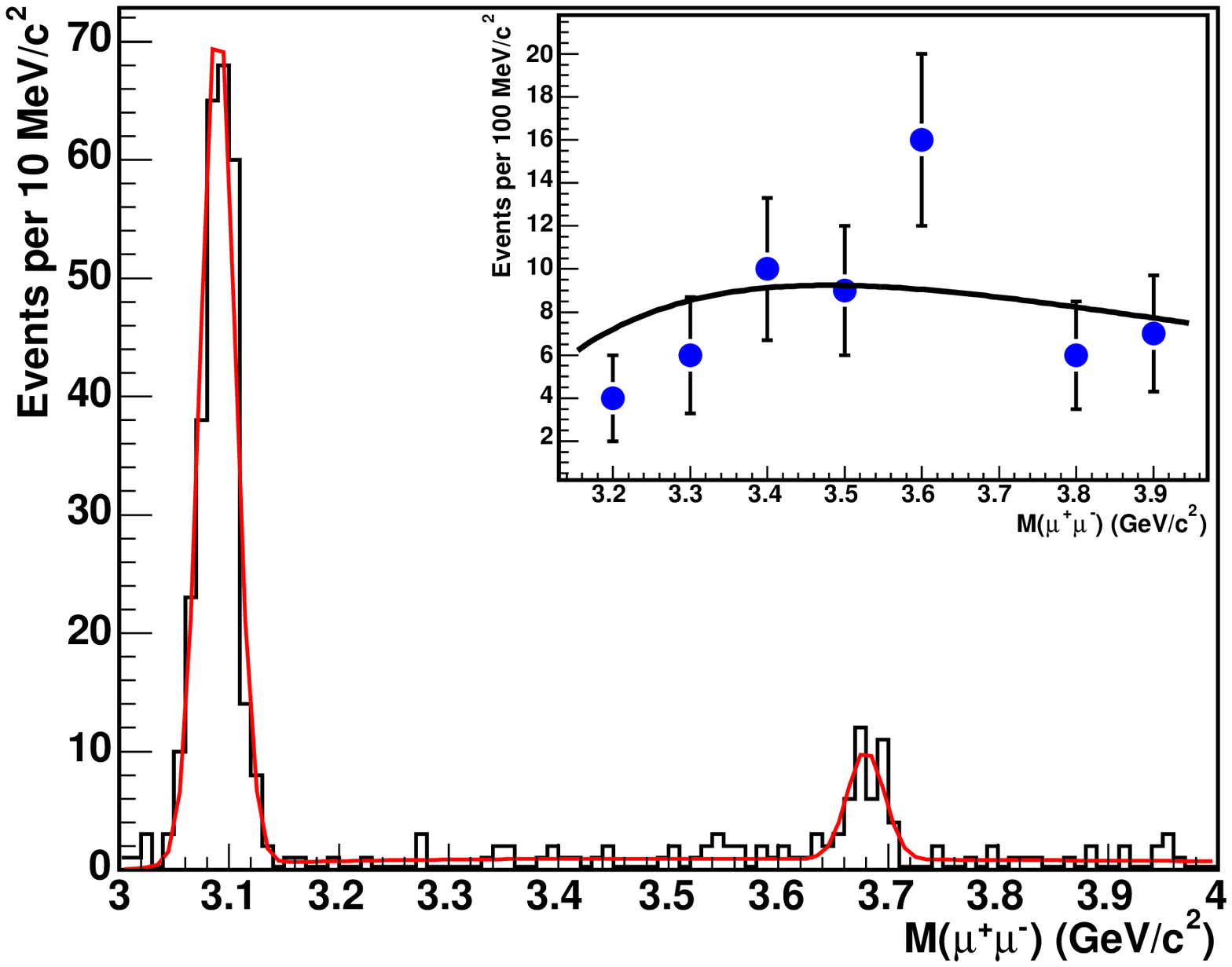}}
}
\vspace*{-0.03\textwidth}
\caption{(a) Transition form factor of $\pi^0$ with latest BaBar measurements (Li); (b) Exclusive dimuon spectrum in CDF, with deiffractive peaks on $\gamma\gamma\to\mu^+\mu^-$ background (Nystrand).
}
\label{Fig:gamgam}
\end{figure}

\section{Two-photon processes}

Selina Li presented some major new photon-photon studies at BaBar
and Belle, reminding us again of the original subject-matter of the
Photon series of conferences!  The process $\gamma\gamma\to
\pi^0\pi^0$ has been accurately measured by Belle for dipion masses
above 0.6 GeV, and BaBar have measured the $\pi^0$ transition form
factor in a single-tagged analysis, matching up with data from CLEO
and also CELLO at DESY's PETRA collider
(Fig.~\ref{Fig:gamgam}(a)). These are just a few topics, and there
remains a rich field of work here to be continued by Belle.

Klaus Dehmelt reminded us of the photon leptonic structure function,
as measured by L3. More results are coming out in this area, in
particular determinations of the structure functions as a function of
both photon virtualities; unsurprisingly, the results are in good
agreement with the QED calculation.  Richard Nisius surveyed the
current state of play with the hadronic photon structure
function. There are many contributions to an overall fit, but most of
the precision points are provided by LEP, especially by OPAL.
Although the low-$x$ reach of the measurements is limited, so that the
predicted low-$x$ rise is not yet experimentally established, a charm
contribution is evident.  We keenly await further results from BaBar and
Belle, which should soon appear.

Two-photon collisions have now been observed at hadron colliders!
Joakim Nystrand showed how CDF have events of the type
$\gamma\gamma\to\mu^+\mu^-$ (Fig.~\ref{Fig:gamgam}(b)) and PHENIX have
$\gamma\gamma\to e^+ e^-$, seen in their studies of diffractive
photoproduction of the $J/\psi$ and $\psi'$. So
here, the background is almost as interesting as the signal.

\begin{figure}[t]
\vspace*{-0.04\textwidth}
\centerline{
\includegraphics*[width=.45\textwidth]{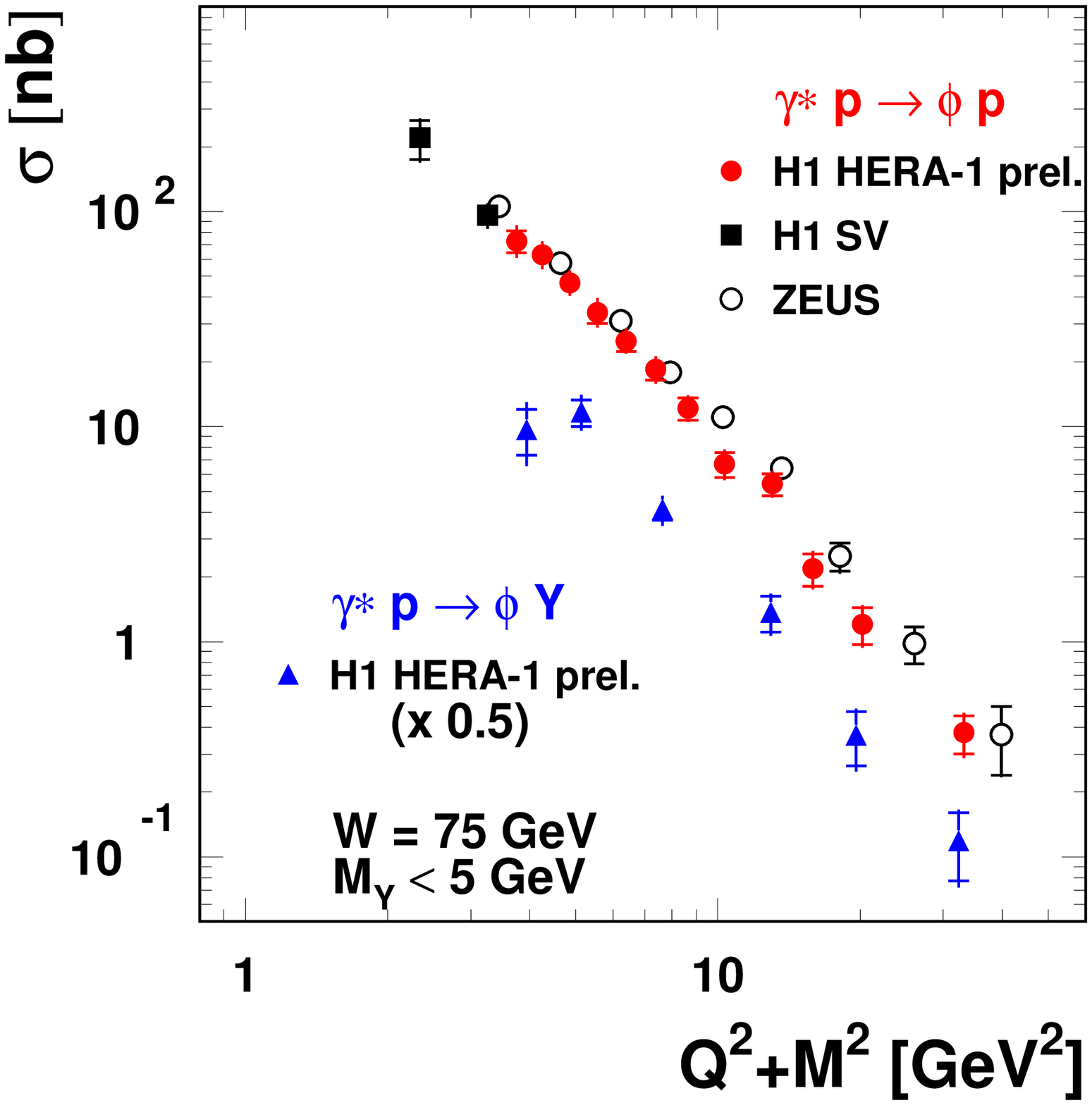}
\raisebox{0.05\textwidth}{
\includegraphics[bbllx=10,bblly=230,bburx=240,bbury=435,width=0.45\textwidth,clip=true]{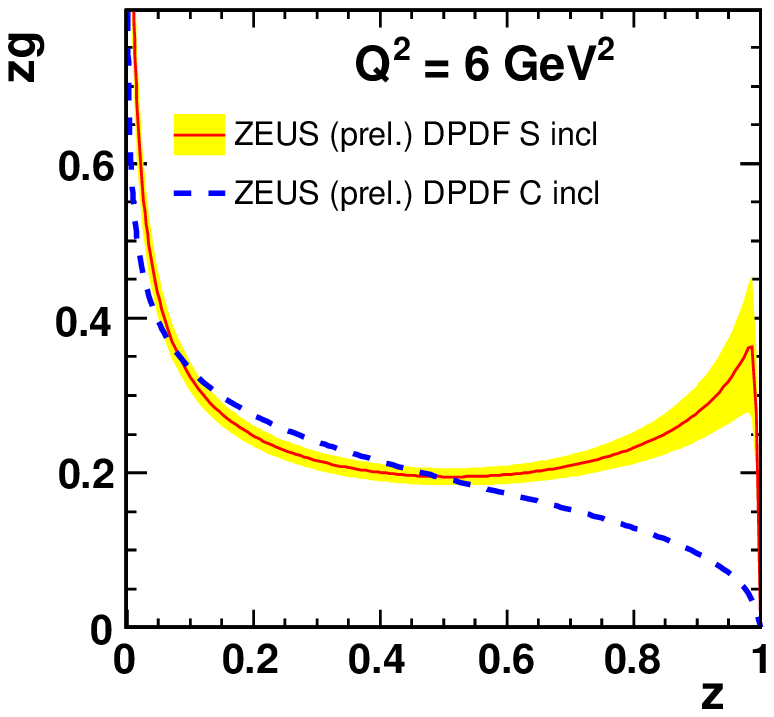}
}
}
\caption{(a) Cross sections for exclusive $\phi$ production at HERA (Kananov); (b) 
Gluon density function in pomeron (Newman).
}
\label{Fig:diff}
\end{figure}

\section{Diffraction}

A number of new results were presented on diffraction in
photoproduction.  At STAR at RHIC, again a hadron collider experiment, photoproduced $\rho$
mesons have been observed and their angular distribution measured.
Andrzej Sandacz showed results from the COMPASS experiment at CERN, using a
160 GeV muon beam on a polarised ammonia target.  In addition to a
strong muoproduction programme, the experiment has measured a variety
of asymmetry parameters in $\rho$ photoproduction from both protons and
deuterons, and have proceeded to extract spin density matrix elements.
Results for $\phi$ production are expected and there are plans for
further measurements concentrating on Deeply Virtual Compton
Scattering (DVCS).

Most of the work in diffractive physics in recent years has come from
HERA.  Sergey Kananov presented a general survey of what we have
learnt on vector meson photoproduction.  There is now an impressive
collection of results from H1 and ZEUS on this subject, and he made a
comparison of vector meson production without and with a hard scale in
the process.  Elastic photoproduction of light mesons, namely the
$\rho$, $\omega$ and $\phi$ shows the normal properties of soft
diffraction, with parameters gently varying with the centre of mass
energy $W$.  The same also holds in electroproduction measured as a
function of $W$. However when a hard scale is present, typified by a
heavy quark or large value of $Q^2$, the cross sections rise with $W$
while falling with $Q^2 + M^2$ (Fig.~\ref{Fig:diff}(a)). These
features are compatible with hard diffraction as evaluated within the
framework of perturbative QCD.
  
One of HERA's major achievements, of course, has been the study of
diffractive physics from the point of view of the pomeron as a
hadronic object with a partonic substructure. Paul Newman presented a
broad overview of the extensive H1 and ZEUS analyses in this area.
Diffractive structure functions have been measured with precision, as
illustrated by recent ZEUS results (Fig.~\ref{Fig:diff}(b)) and a
variety of detailed ideas can be tested, such as the factorisation
properties of the proton vertex as the diffractive process at higher
$Q^2$ looks more and more like a kind of hard gluon exchange.
Different analysis approaches give consistent results, and a pure DIS
or rapidity-gap approach can be successfully compared with the ZEUS
data with forward proton or neutron detection.  H1 have presented the
first diffractive $F_L$ determination.  For the future, the full HERA
II data need to be analysed, and the H1 and ZEUS data combined for
overall measurements.

\begin{figure}[t]
\centerline{
\includegraphics*[width=.45\textwidth]{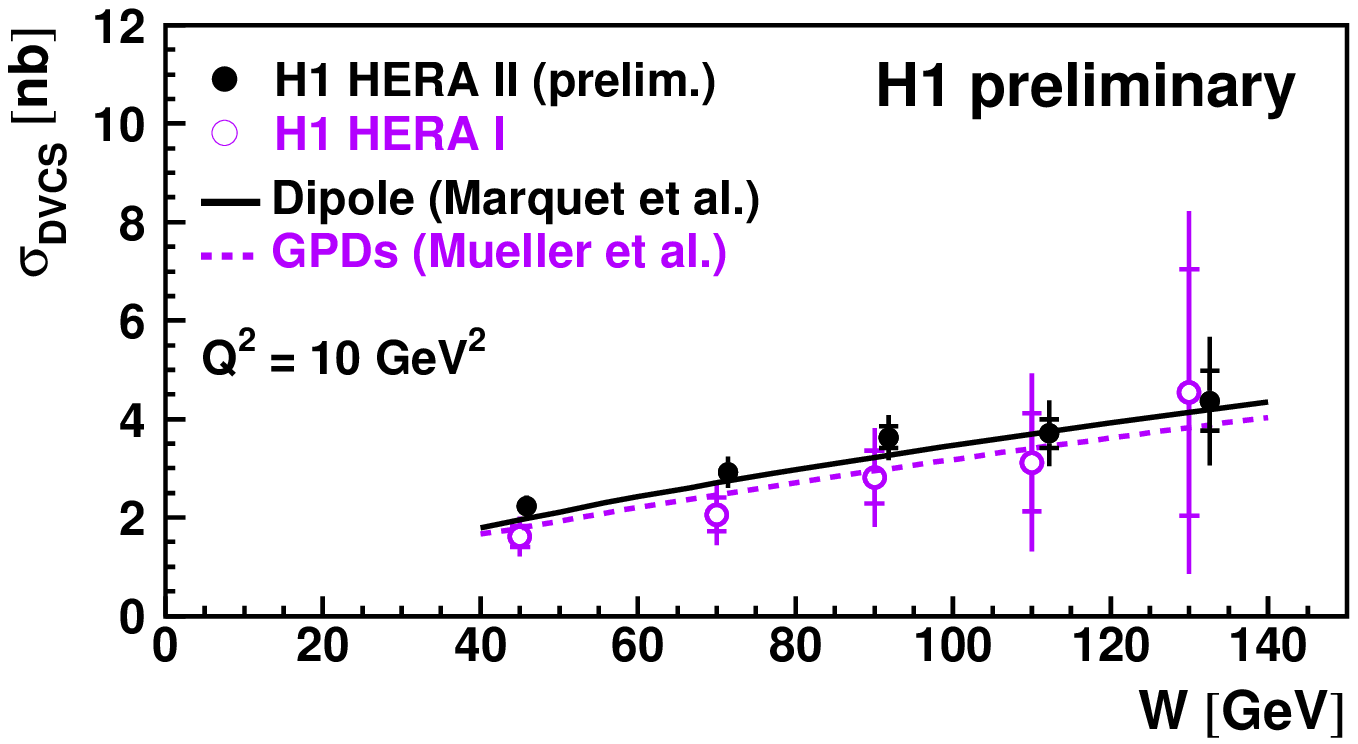}
\raisebox{0.00\textwidth}{
\includegraphics[width=0.45\textwidth,clip=true]{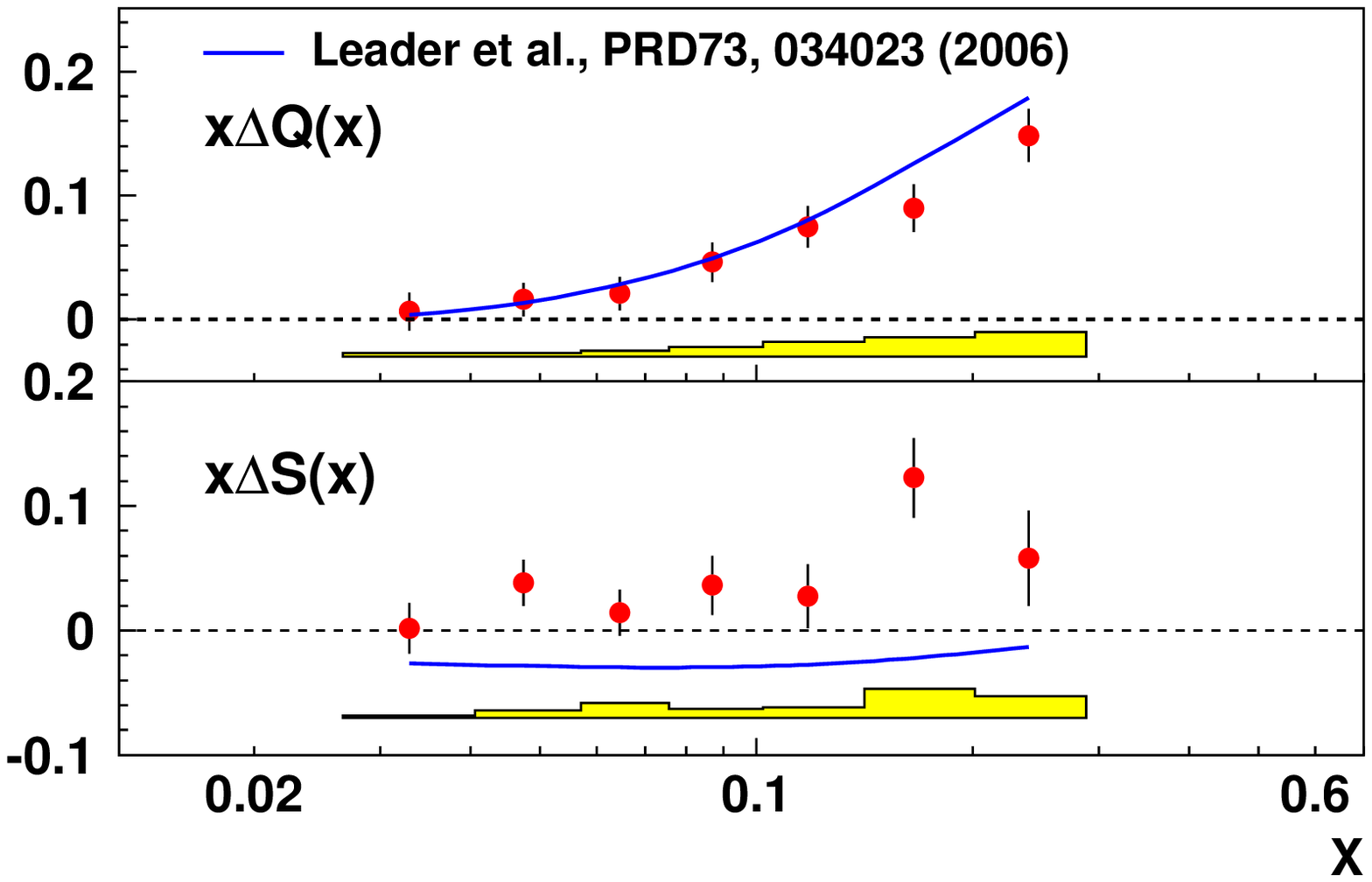}
}
}
\caption{(a) DVCS cross sections at H1 (Schoeffel); (b) 
Nucleon quark helicity distributions at HERMES (Hillenbrand).
}
\label{Fig:dvcs}
\end{figure}

\section{DVCS and proton structure}

Several presentation of results in Deeply Inelastic Compton Scattering
were presented, a topic that requires fairly high integrated
luminosities in order to obtain useful statistics.  Photons are
scattering off protons -- this is high energy photon microscopy!
Laurent Schoeffel described results from H1 and ZEUS.  At these low $x$
values, for $x<0.01$ in the proton, the relevant gluon
density is high and one might need to think about saturation
effects. Direct DVCS can be measured and the $Q^2$ and $W$ dependence
analysed.  However there is an irreducible Bethe-Heitler background of
a similar magnitude.  Both experiments have presented new cross
sections and H1 show that both a dipole model and a Generalised PDF
model can fit the data.  Are GPDFs as opposed to simple PDF's needed?
H1 present clear evidence that argues for a skewing effect, in support
of the idea of GPDFs. In their upgrade, COMPASS are making DVCS a
major focus and will continue these investigation.  Measurement of the
Beam Charge Asymmetry will provide an important tool for the more
detailed studies, and first measurements of this quantity have already
been made by H1.

Frank Sabati\'e gave a dedicated talk on DVCS measurements at JLAB.
High statistics are available making use of intense electron beams,
enabling several angular correlations to be measured and asymmetries
to be determined, together with detailed determinations of the
amplitude properties evaluated from suitable cross section
differences.  It is found that the data are badly described by several
of the simpler models, suggesting the presence of more complex or
higher order effects.

A further perspective on DVCS was given by Achim Hillenbrand in a
presentation of a variety of highlights from the HERMES experiment.
HERMES have measured many aspects of nucleon structure using hydrogen
targets as well as different atomic nuclei.  A particular emphasis has
been on spin structures (Fig.~\ref{Fig:dvcs}(b)), a topic also taken
up by Joerg Pretz in a discussion of the helicity contribution of
gluons to the proton spin structure, measured by means of charmed
meson distributions in the COMPASS experiment.  Max Klein discussed
one of HERA's major showpieces, the DIS study of the PDFs of the
proton, using combined H1 and ZEUS data.  A new fit has been made to
these results, confirming the global utility of perturbative QCD, and
comparisons are being made to data from the Tevatron. Specific proton
structure functions are being evaluated for charm and beauty
production.  More will come at higher values of $Q^2$ and $y$ as well
as for the heavy flavours, together with more on $F_L$, which has now
been measured by H1 and ZEUS.  It can be said that quarks are
pointlike down to $0.7\times10^{-18}$m, making HERA also the world's best
electron microscope.

\begin{figure}[t]
\centerline{
\vspace*{-0.007\textwidth}
\includegraphics*[bbllx=0,bblly=0,bburx=543,bbury=621,width=0.45\textwidth,clip=true,width=.5\textwidth]{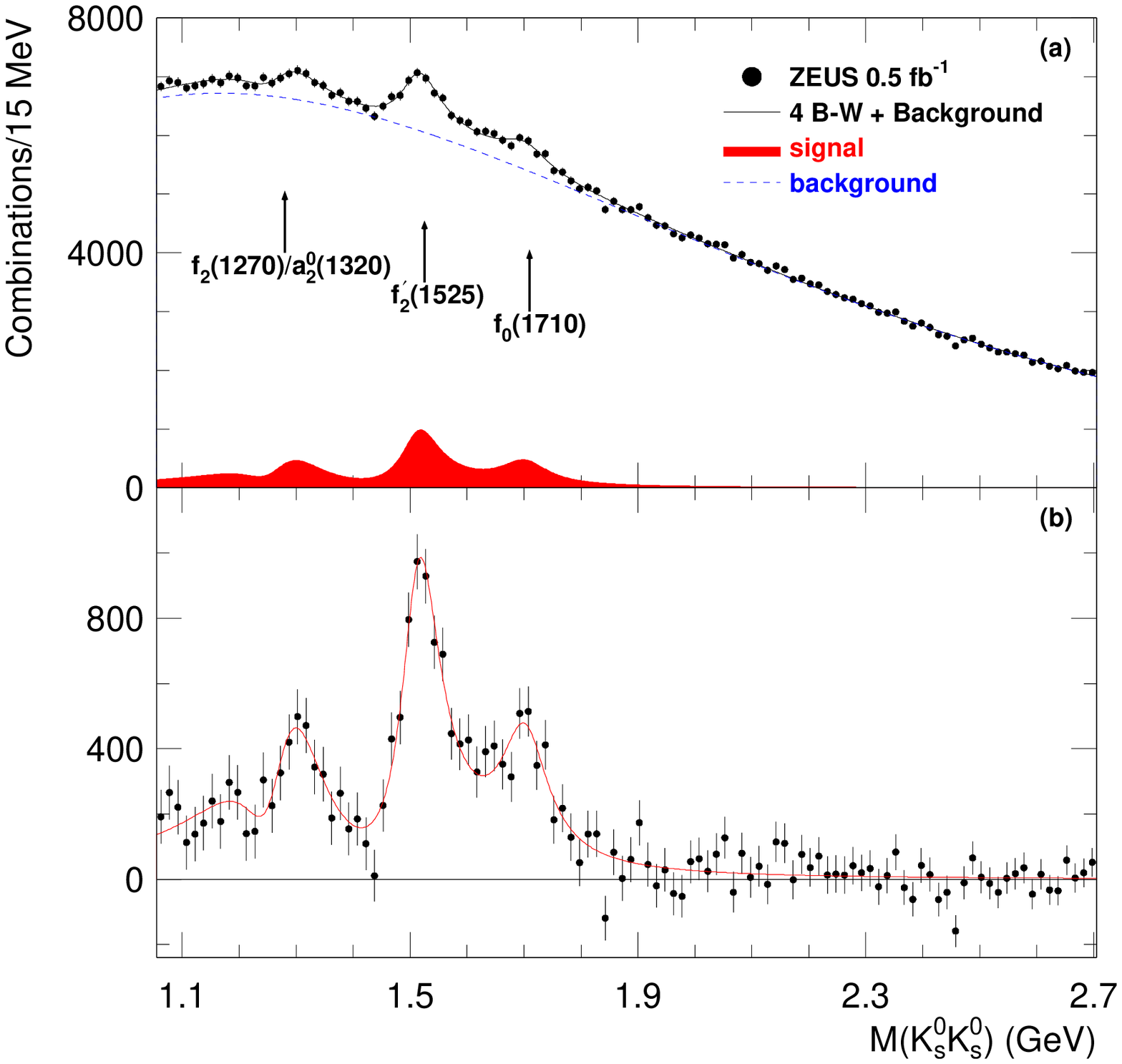}
\raisebox{0.40\textwidth}{
\includegraphics[bbllx=0,bblly=260,bburx=415,bbury=500,width=0.45\textwidth,clip=true,width=.45\textwidth]{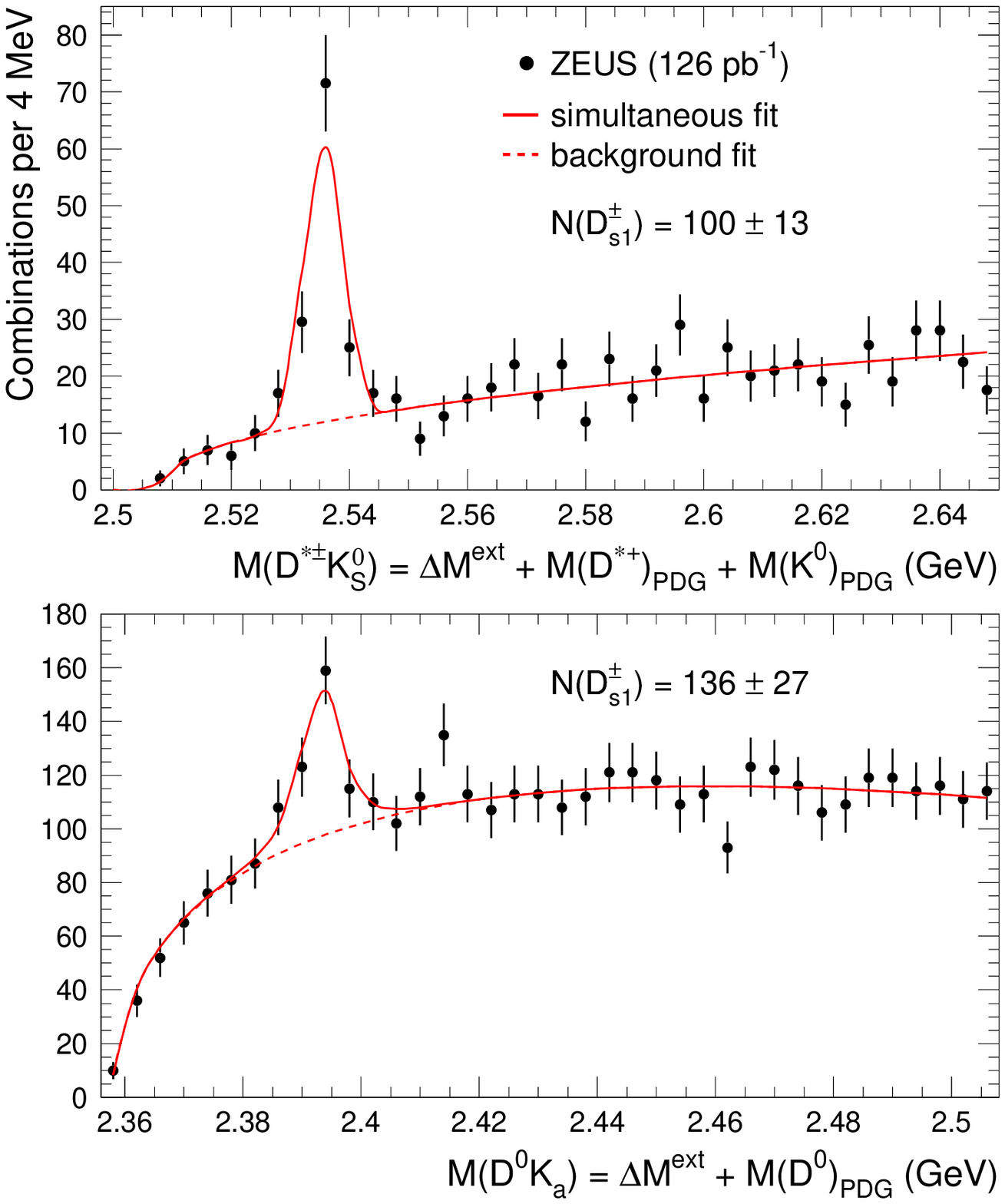}
}
}
\vspace*{-0.11\textwidth}
\caption{(a) Structure in $K^0K^0$ spectrum at ZEUS; (b) 
$D(2536)$ signal in mass difference spectrum (Karshon).
}
\label{Fig:res}
\end{figure}

\section{Resonance production}

As a high energy collider, HERA produced final states
containing the usual variety of hadronic resonances. This has provided the
opportunity to search for new or exotic resonances that had in some
cases been reported at other colliders.  Uri Karshon presented results
on several such searches,one being for glueballs in the
$K^0_sK^0_s$ system.  The scalar meson sector contains too many 0++
states to fit into the normal quark model, and it is natural to
investigate these as possible glueballs, the lightest of which is
predicted by lattice gauge calculations to have a mass in the range
1550-1750 MeV. The state $f_0(1710)$ is an interesting glueball
candidate, for several reasons including an apparently high decay BR
into strange quarks. The $K^0_sK^0_s$ system is therefore an
attractive area to look for confirmation of these ideas. ZEUS have
observed evidence for several $K^0_sK^0_s$ resonances in this mass
region and the $f_0(1710)$ is present with good statistical
significance. It is argued, however, that the state is not a pure
glueball.

H1 had earlier proposed a charm pentaquark signal at 3.1 GeV.  However
ZEUS did not see this, and with more HERA II statistics this peak has
gone away and presumably must be treated as a statistical fluctuation,
albeit a rather enigmatic one.

ZEUS are studying excited charm states, which are well observed even
in the HERA I data set.  A variety of states are observed, and 
the availability of the HERA II data with a better vertex detector 
makes this a very promising prospect for the future.  

\begin{figure}[t]
\centerline{
\includegraphics*[width=.45\textwidth]{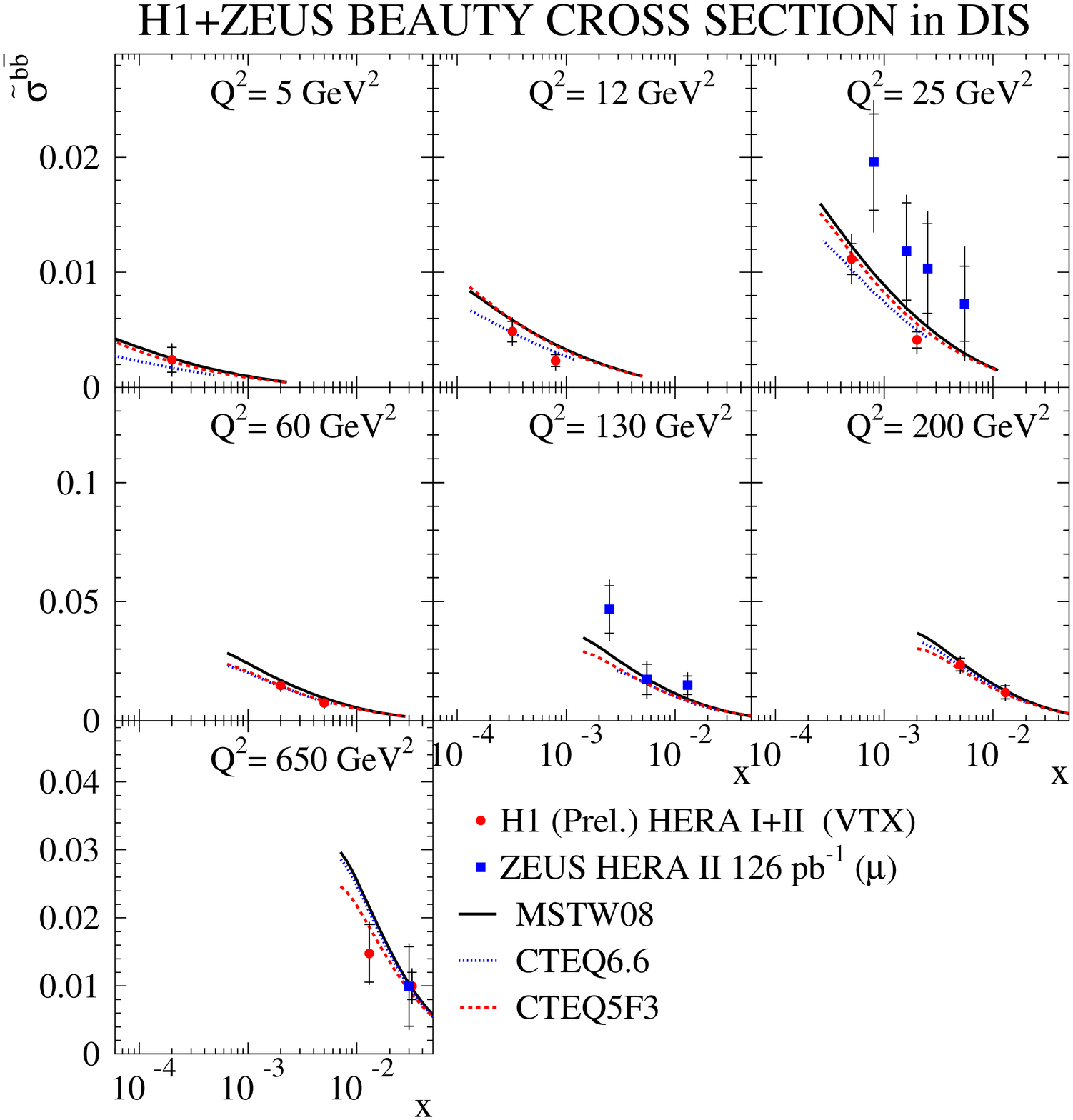}
\raisebox{0.00\textwidth}{
\includegraphics*[width=0.45\textwidth,clip=true,width=.45\textwidth]{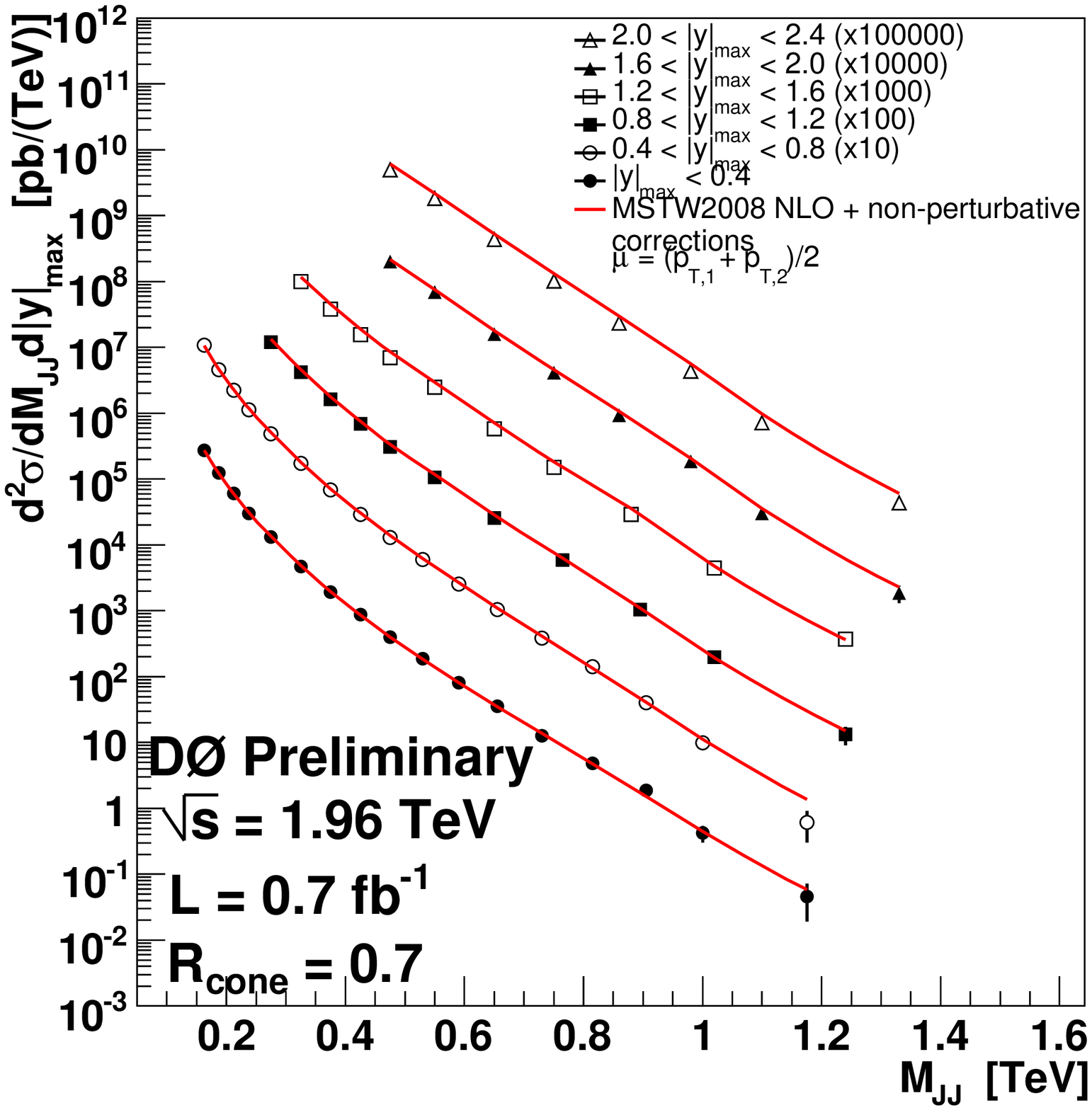}
}
}
\caption{(a) Combined HERA beauty DIS data (Grindhammer); (b) 
D0 dijet mass spectrum (Yu).
}
\label{Fig:jets}
\end{figure}

\section{Jet physics}

The theme of heavy flavours formed a major aspect of G\"unter
Grindhammer's discussion of the physics at different hard scales.
Both in photoproduction and in DIS, the presence of a heavy quark
facilitates the use of perturbative QCD and encourages the calculation
of theoretical models. It is found that the HVQDIS model describes
charm production in DIS well, but predicts too low cross section for
beauty.  Overall, however, NLO and NNLO calculations do a good job at
describing the features of the data (Fig.~\ref{Fig:jets}(a)). Results
from the final HERA data sets are very eagerly anticipated. In another
part of this talk, the evaluation of $\alpha_S$ from jet measurements
at HERA was surveyed.  H1 and ZEUS are able to do this in various ways
and many of the measurements are very competitive on a world basis.
It remains the case that with the increasing experimental precision,
theoretical uncertainties dominate most of these determinations.

The story is continued with the study of jets at the Tevatron, presented
by Shin-Shan Yu.  Both CDF and D0 have accumulated in their data very
large samples of jets, produced up to transverse momentum values above
600 GeV/$c$.  At one time, high-$p_T$ anomalies were suggested but at present
everything is well described by the latest fits to the proton structure which are
based on lower-$p_T$ data.  Everything about inclusive jets and dijets
currently looks very satisfactory (Fig.~\ref{Fig:jets}(b)).  Studies of $W$ and $Z$
accompanied by jets and, specifically, by heavy flavours have been
carried out.  These will be of particular relevance in connection with
Higgs searches.  Anne-Marie Magnan took the view to a higher energy level,
presenting projected jet features at LHC.  Not only are the extensive
studies being performed here important with regard to understanding
QCD and the proton structure in more depth, they will be esssential in
understanding the backgrounds to searches for exotics and Higgs at LHC.

\begin{figure}[t]
\centerline{
\raisebox{0.385\textwidth}{
\includegraphics[bbllx=160,bblly=25,bburx=490,bbury=530,width=0.38\textwidth,angle=270,clip=true]{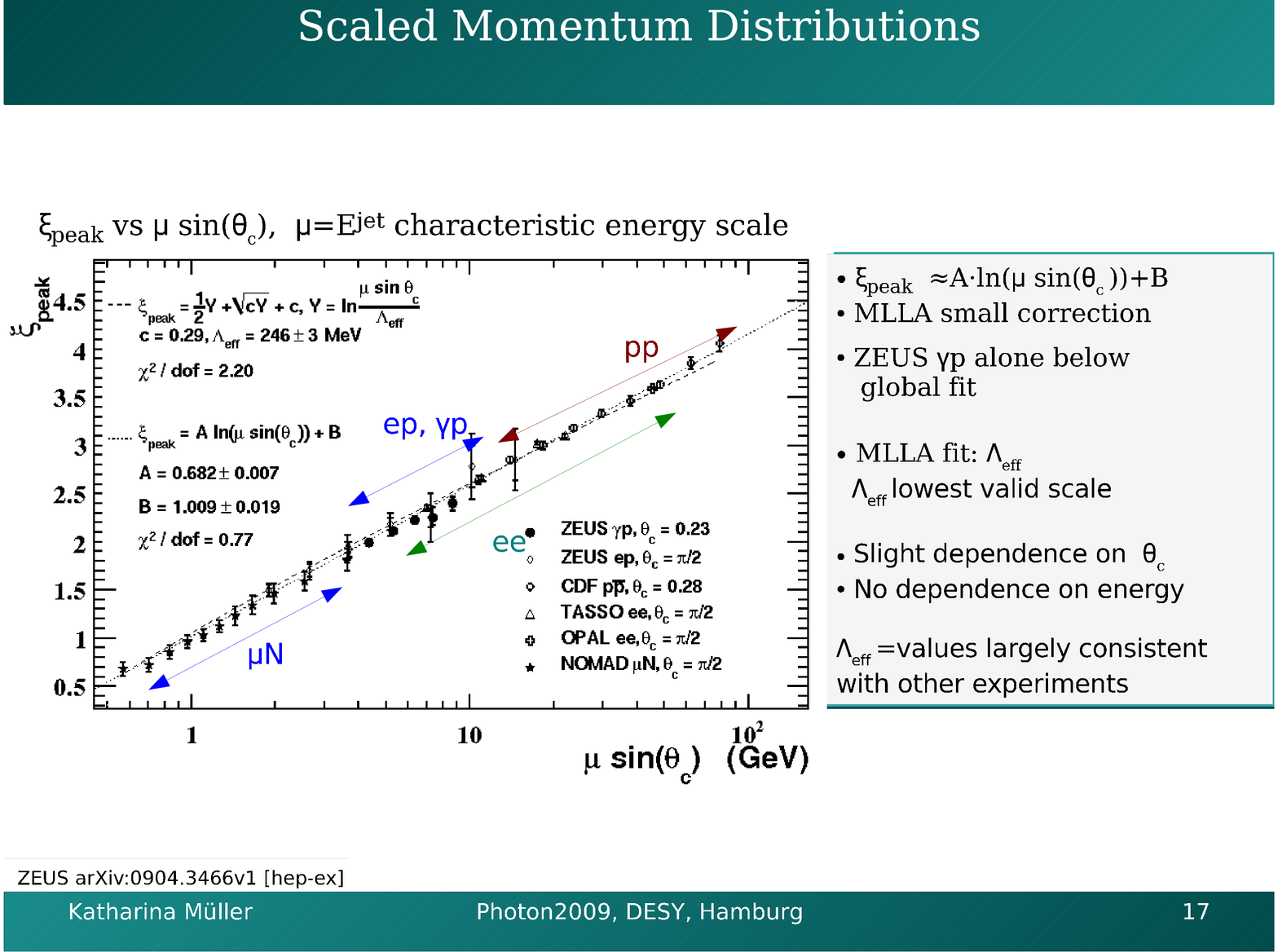}
}
\includegraphics*[width=0.45\textwidth,width=.45\textwidth]{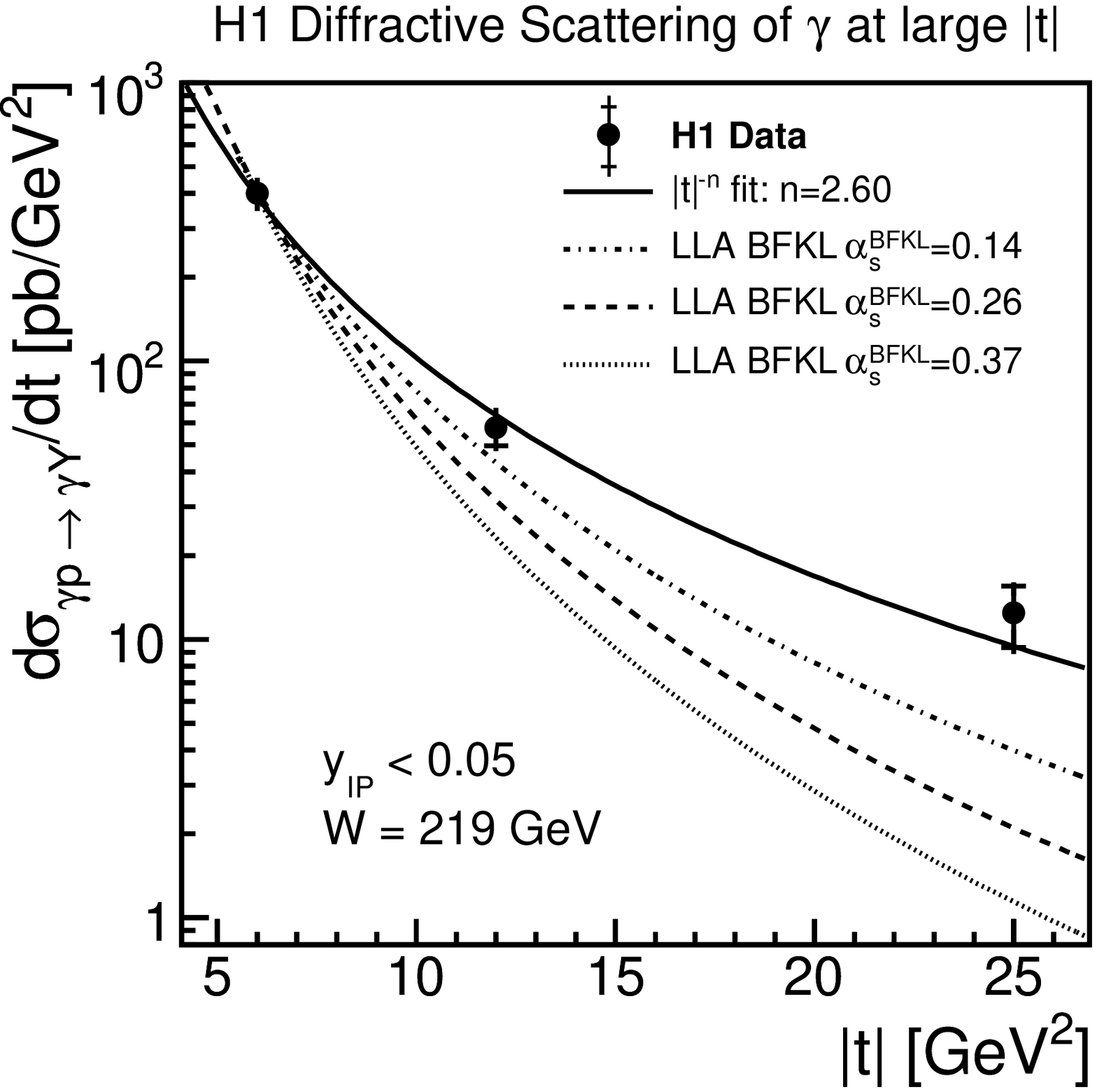}
}
\caption{(a) Universal properties of scaled momentum in fragmentation; (b) 
high-$t$ photon distribution (Mu\"ller).
}
\label{Fig:photo}
\end{figure}

\section{Photoproduction}
Most of the events recorded in the two larger HERA experiments
lie in the category of photoproduction, in which the virtuality of the
exchanged proton is very much lower than 1 GeV$^2$.  A thorough survey
of many aspects of these processes was given by Katherina
M\"uller. Hard photoproduction may be specified as comprising
processes in which either the photon itself or a parton within it
undergoes hard scattering, commonly giving rise to jets in the final
state. The photon in this way often behaves as if it has a hadronic parton structure. 
Inclusive jets, dijets and the properties of dijets have been measured and show no
unusual properties; neither does the topology of the jets nor the
fraction of the photon momentum that is taken up in the jet
formation. 

\begin{figure}[t]
\centerline{
\includegraphics*[width=.5\textwidth]{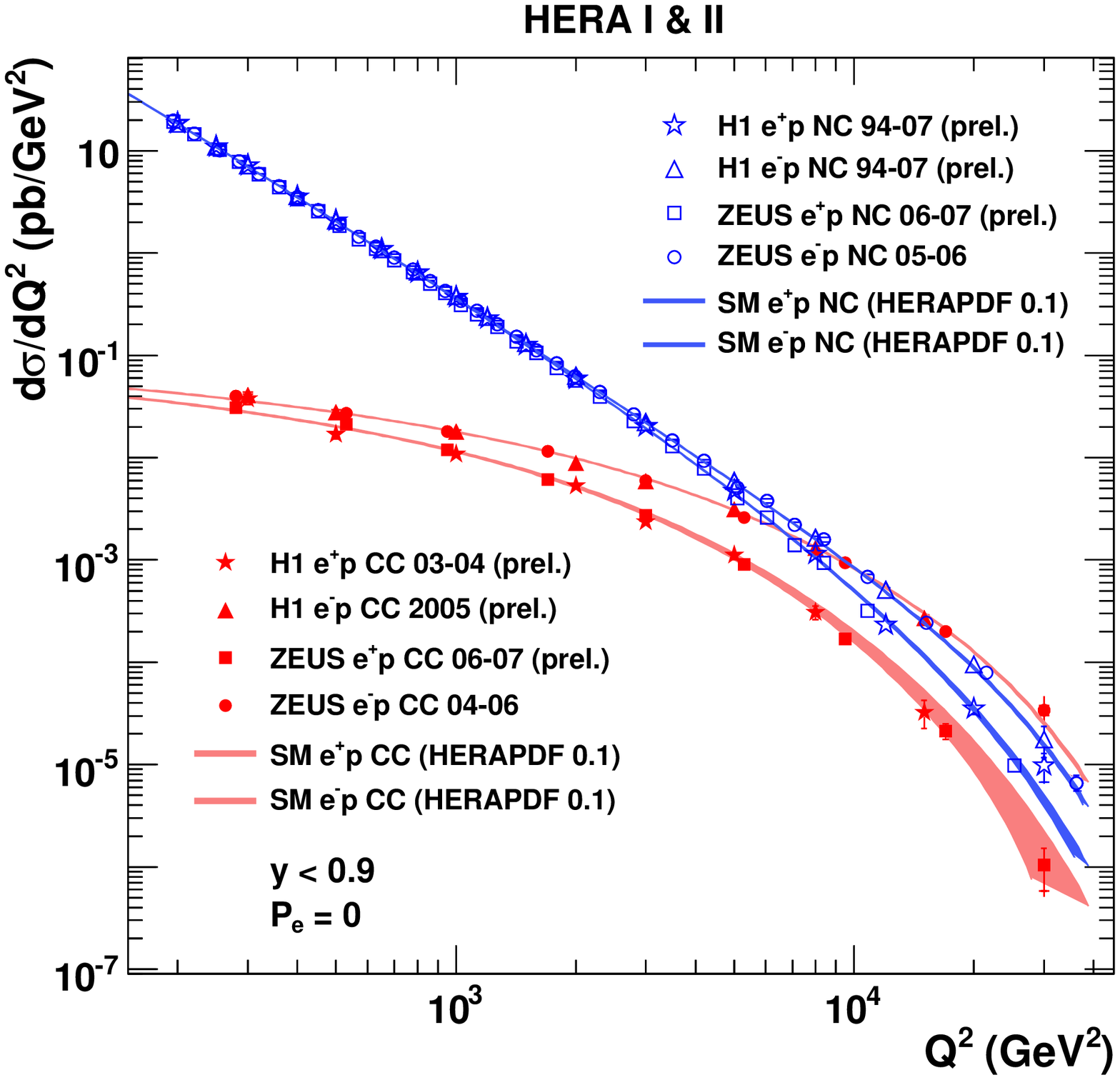}
\raisebox{-0.03\textwidth}{
\includegraphics*[width=0.45\textwidth,width=.49\textwidth]{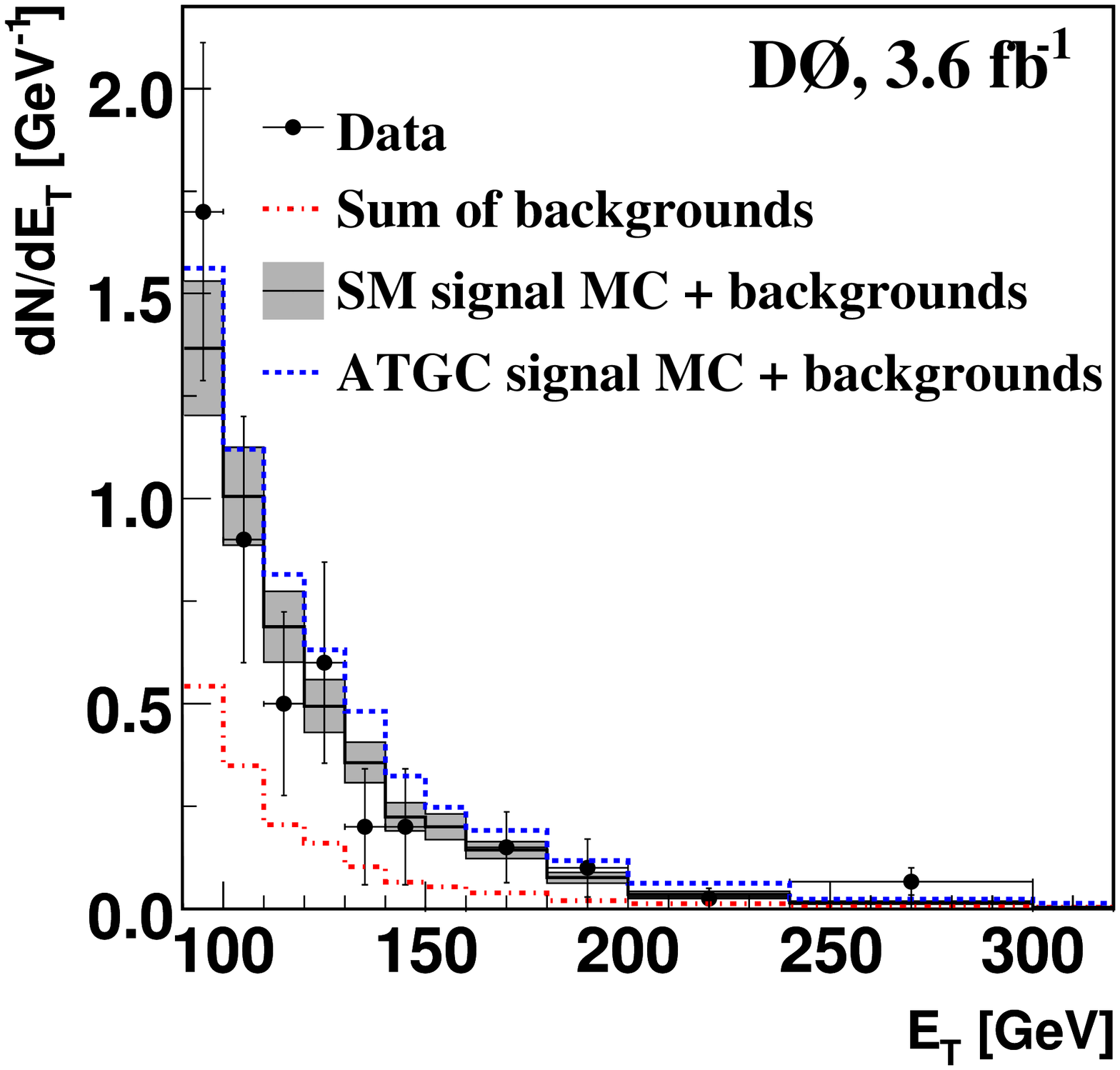}
}
}
\caption{(a) H1 + ZEUS electroweak DIS cross sections (Stanco); (b) 
D0 data showing no evidence for anomalous triple gauge couplings (Krop).
}
\label{Fig:high}
\end{figure}

In a recent analysis, ZEUS have given measurements of scaled momentum
distributions in photoproduced jets, testing our understanding of
fragmentation and its universality when this is compared with
appropriately scaled measurements from other regimes; again, all is in
order (Fig.~\ref{Fig:high}(a)). However the diffractive scattering of
high-$t$ photons was found by H1 to be higher than expected and topic
well merits further study (Fig.~\ref{Fig:photo}(b)).

\begin{figure}[t]
\centerline{
\includegraphics*[width=.52\textwidth]{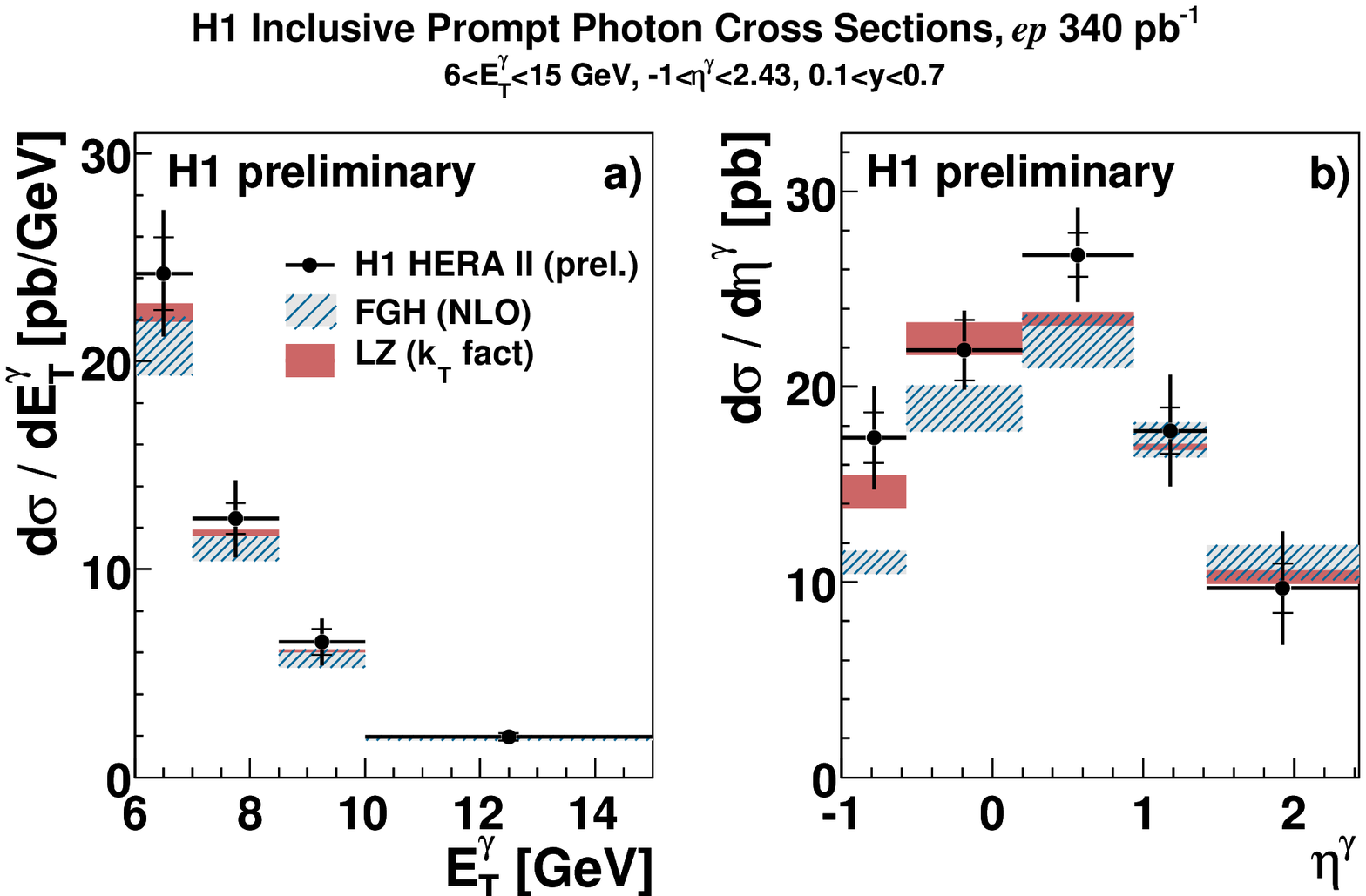}
\raisebox{0.00\textwidth}{
\includegraphics*[width=0.47\textwidth,width=.5\textwidth]{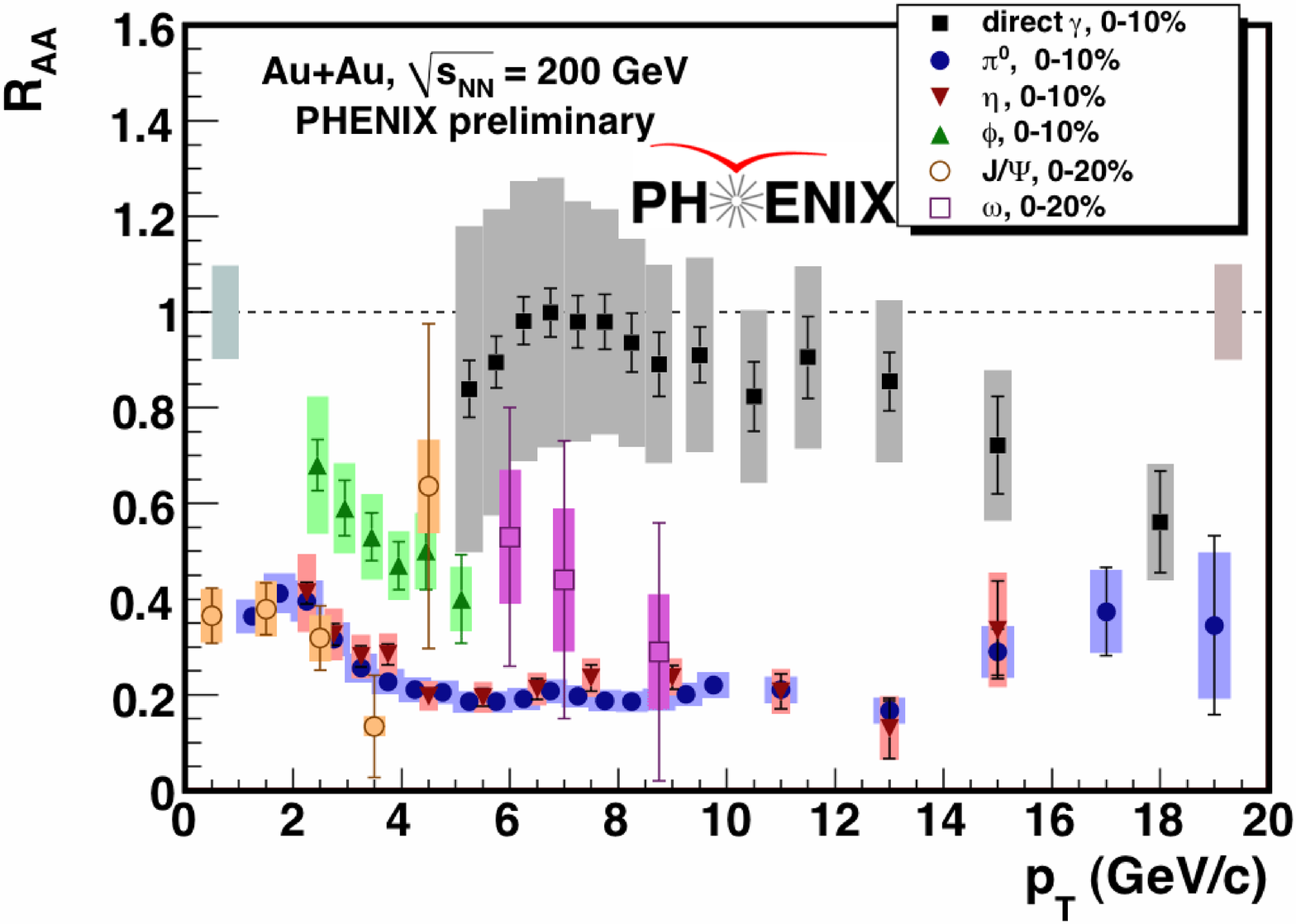}
\hspace*{-0.05\textwidth}
}
}
\caption{(a) H1 prompt photon + jet in photoproduction (M\"uller)); (b) 
PHENIX enhanced prompt photon ratio, compared to other emitted particles (Reygers).
}
\label{Fig:prp}
\end{figure}

\section{Prompt photons}
In the study of high energy collisions involving hadrons, events in
which an isolated high-energy photon is observed provide a direct
probe of the underlying parton process, since the emission of these
photons is largely unaffected by parton hadronisation.  The study of
such ``prompt'' photons gives new perspectives on QCD processes,
allowing theory to be tested from new viewpoints.  Prompt photons may
be emitted in hard partonic interactions, and were the subject of part
of the talk on photoproduction. H1 have measured prompt photons in
photoproduction accompanied by a jet, giving cross sections and
distributions in the azimuthal separation of the photon and the jet(Fig.~\ref{Fig:prp}(a).
These distributions are compared with models of the photon, with fair
but not perfect agreement.

Recent results on prompt photons in DIS in ZEUS were presented by
myself.  In this process, attention must be paid to the radiation of
the photon by either the incoming or outgoing lepton.  Agreement with
theoretical models is fair but shows some serious disagreements in
some kinematic regions.

At the Tevatron, the story was continued by Ashish Kumar.  Both CDF
and D0 are active in the prompt photon area.  This process is sensitive
to the structure of the proton, as well as to the possibility of new
physics.  CDF and D0 find good agreement with theory, to within some
fairly substantial theoretical uncertainties, for prompt photons at
transverse momenta above 50 GeV.  Below this CDF see a discrepancy
while with D0 the situation is suggestive but a little
ambiguous -- this region clearly merits further study.  Measuring a jet
as well as the photon does not bring the situation fully under
control, as measured by D0. Demanding a $b$-jet does achieve agreement
with theory, but there are serious disagreements if the photon is
accompanied by a $c$-jet.  Again, more study is indicated.

Prompt photons also form an important topic of investigation at the
RHIC collider, as reported by Klaus Reygers. The PHENIX results in
proton-proton collisions are consistent with a large collection of
results from other colliders.  However, in nucleus-nucleus collisions,
it is understood that the colliding nuclei are likely to form a
quark-gluon plasma or fireball, out of which prompt photons can
emerge.  At photon energies of a few GeV this effect has apparently been 
observed by PHENIX using gold-gold collisions.  There is indeed an
enhancement of these photons compared to the distributions observed in
proton-proton collisions (Fig.~\ref{Fig:prp}(b)), confirming the idea
that a quark-gluon plasma is being formed.

\section{The present high energy frontier}
As energies rise, we enter the realm of electroweak physics, and of
many possibilities of new physics.  Luca Stanco presented some updated
HERA cross sections comparing neutral current and charged current
exchange.  The already classic HERA I results are now augmented by
preliminary HERA II measurements, giving further accuracy, and we see
how pure photon exchange merges into $Z$ exchange at the same level as
$W$ exchange, with differences between electron and positron cross
sections (Fig.~\ref{Fig:high}(a)). The beam polarisation asymmetries
illustrate the physics equally powerfully.

At the Tevatron, Dan Krop pointed out the plethora of theoretically
proposed new processes that involve photons.  These range from the
familiar area of Higgs physics, through SUSY models and extra
dimensions, to Compositeness, new generation(s) and Technicolor.  All
these areas are the subject of searches at the Tevatron and all the
searches have so far proved unsuccessful. The anomalous coupling of
the photon to the $Z$ is one such example (Fig.~\ref{Fig:high}(b)),
although the search has given the first observation (by D0) of the
production of $Z\gamma\to\nu\nu\gamma$ at the Tevatron. There has also
been an unsuccessful search for ``dark photons'' that might explain
certain excesses seen in astroparticle physics experiments.  No Higgs
signal, which would be non-SM under present conditions in the
photon-photon channel, has been observed.  There are many other
possibilities, limited only by the imagination of theorists, for
anomalous physics to be observed in conjunction with photons.  These
searches will continue until the Tevatron terminates and LHC takes
over the baton.

At LHC, as explained by Suen Hou, there will be a lot of work to do in connection with photons.
There are many Standard Model processes that involve high energy
photons, all of which should be studied.  Some of these are standard
processes, such as $W$ and $Z$ production, in which a photon is
radiated by an incoming quark line: a correct understanding of this
kind of process will help ensure that we understand the basic $W$ and
$Z$ production correctly. Of course, the search for anomalous
couplings will be extended.  The decay of the Higgs into two photons
will be a major focus of LHC work, as elaborated by David Joffe in a
talk that presented much information on the techniques of photon
detection at ATLAS and CMS.  While we are waiting for the Higgs to be
discovered, it will be interesting to measure the proton-proton total cross section
(Hasko Stenzel).

Finally, there are interesting prospects for photoproduction physics at the 
LHC, since the high energy protons are quite efficient at radiating photons.
Vincent Lema\^itre outlined this photoproduction programme, concentrating
on the possibility of single top production via an incoming photon.  The process
will be tagged, it is hoped, by installing forward silicon detectors close
to the beamline downstream of the detectors.  Nicolas Schul extended this discussion
to a programme of photon-photon physics at the LHC, epitomised by SUSY searches
which can be performed in a very effective way using this approach. 

\section{A high energy photon collider?}

For the far future, plans have been under development for many years
to construct a high energy photon photon collider as a part of the
International Linear Collider project.  Talks by Valery Telnov and
Tohru Takahashi presented some technical ideas that might be able to turn this
project from aspiration into reality, while Jeff Gronberg discussed
the possible benefits of constructing an ``early photon collider''
that could, for example, become a kind of Higgs factory by
manufacturing the Higgs partons out of pairs of photons, the converse
of the decay process that will be eagerly sought at the LHC.  Unfortunately,
these are financially very difficult times, and the present climate of
opinion is unfavourable to the pursuance of this option.

\section{Final remarks}

In this overview I have attempted to give an impression of the
remarkable range of talks given at the Photon 2009
conference. There is an enormous diversity of particle physics
processes in which photons, whether incoming or outgoing, real or
virtual, play an important role.  The fact that a particle is ubiquitous
does not make it less interesting than those that are rarer; quite on the
contrary, the humble photon provides a key to the deeper
understanding of many things.  From low energies to high, from the
firmly established to the speculatively hypothetical, we have seen how
important photon physics is to all aspects of our subject. It seems
that there is no other particle in the universe that serves us in so
many ways.

I must express regrets with respect to those topics which this review
has had to omit. These include the DESY laboratory's extensive
programme of ``low-energy'' photon research, and the entire area of
new developments in photon detectors.  I have also not been able to
cover the subject of photons in astroparticle physics.  These topics
are treated in their own papers in this volume.

Above all, I would like to thank the organisers of the conference for
their dedication in making possible such an excellent week of presentations 
and discussions, in which an outstanding breadth of fascinating
physics was presented.  We all look forward keenly to the next
conference in the Photon series.

\end{document}